\documentstyle[fancyhdr,aps,epsfig,epsf,eqsecnum,multicol,prb]{revtex}

\def\breakon{\end{multicols}\widetext\vskip -3.0cm\begin{picture}(290,80)(80,500)
\thinlines
\put(65,500){\line( 1, 0){255}}\put(320,500){\line( 0, 1){  5}}\end{picture}}
\def\breakoff{\vskip -3.0cm\begin{picture}(290,80)(80,500)\thinlines \put(330,500)
{\line( 1, 0){255}}\put(330,500){\line( 0, -1){5}}\end{picture}
\begin{multicols}{2}
\narrowtext}
\newcommand{\be}{\begin{equation}}
\newcommand{\bea}{\begin{eqnarray}}
\newcommand{\ba}{\begin{array}}
\newcommand{\ee}{\end{equation}}
\newcommand{\eea}{\end{eqnarray}}
\newcommand{\ea}{\end{array}}
\newcommand{\nn}{\nonumber}

\renewcommand{\vec}[1]{\mbox{\boldmath$#1$}}
\newcommand{\bm}{\boldmath}

\newcommand{\tr}{{\rm tr}}

\begin {document}
\draft
\rhead{\small cond-mat/}
\chead{}
\lhead{\small }
\lfoot{}
\rfoot{}
\cfoot{\thepage}

\title{Magnetic impurities in the one-dimensional spin-orbital model}
\author{Yu-Wen Lee and Yu-Li Lee}
\address{Physics Department, National Tsing Hua University, Hsinchu, Taiwan}
\maketitle
\begin{abstract}
Using one-dimensional spin-orbital model as a typical example of quantum spin systems
with richer symmetries, we study the effect of an isolated impurity on its low energy 
dynamics in the gapless phase through bosonization and renormalization group methods.
In the case of internal impurities, depending on the symmetry, the boundary fixed points
can be either an open chain with a residual spin or (and) orbital triplet left behind, or a periodic chain. However, these two fixed points are indistinguishable in the sense that
in both cases, the {\it lead-correction-to-scaling boundary operators} (LCBO) only show Fermi-liquid like corrections to thermodynamical quantities. (Except the possible 
Curie-like contributions from the residual moments in the latter cases.)
In the case of external (Kondo) impurities, the boundary fixed points,
depending on the sign of orbital couplings, can be either an open chain with an isolated orbital doublet due to Kondo screening or it will flow to an intermediate 
fixed point with the same LCBO as that of the {\it two-channel} Kondo problem.
Comparison with the Kondo effect in one-dimensional (1D) Heisenberg spin chain and multi-band Hubbard models is also made.
\end{abstract}

\begin{multicols}{2}

\section{introduction}

In the past few years, there have been intensive studies on the one-dimensional spin-orbital model both analytically\cite{azaria1,affleck2,azaria2} and numerically
\cite{udea1,mila,udea2}. Part of the reasons stems from the belief that the unusual 
magnetic 
properties observed in some recently discovered qusai-one-dimensional spin gapped 
materials such as Na$_2$Ti$_2$Sb$_2$O\cite{axtell} and NaV$_2$O$_5$\cite{udea3} 
can be 
explained by a simple  two-band Hubbard model at quarter filling. Owing to the strong
Coulomb repulsion, the corresponding low energy effective Hamiltonian 
can then be mapped onto a quantum spin model:
\bea
H&=&\sum_i J_1  \ {\vec S}_i \cdot{\vec S}_{i+1}+J_2 \ {\vec T}_i \cdot{\vec T}_{i+1}
\nonumber\\
&&+~K \ ( {\vec S}_i\cdot{\vec S}_{i+1} )( {\vec T}_i\cdot{\vec T}_{i+1} )  \ ,
\label{ham1}
\eea
where ${\vec S}_i$ and ${\vec T}_i$ are spin one-half operators representing spin and 
orbital 
degrees of freedom at each site, respectively.
For generic couplings $J_{1(2)}$ and $K$, the model (\ref{ham1}) has a SU(2)$_s$
$\otimes$ SU(2)$_t$ symmetry. However, at the special couplings $J_1=J_2=K/4$,
the symmetry group is enlarged to SU(4), which is Bethe ansatz integrable
\cite{sutherland}. The low energy effective theory at this point is known to be described
by the SU(4)$_1$ Wess-Zumino-Novikov-Witten (WZNW) model, with central charge $c=3$
, equivalent to three decoupled free bosons\cite{affleck3}.

Besides being as a quantum spin model for quasi-one-dimensional materials, the 
spin-orbital model can also appear as  a low energy effective theory in other context. An
example of this is the spin-tube model  studied recently by E.Origanc et al.\cite{andrei}.
In that case, the spin and orbital operators do not represent the real spin or orbital degrees
of freedom, but are just mathematical objects used to describe the degenerate ground states
obtained after projecting  out the high energy states in the Hilbert space via 
renormalization group transformation.

Earlier studies of the model (\ref{ham1}) around the SU(4) point were concentrated
on the Z$_2$ symmetric case $J_1=J_2=J$. The results show that when $J<K/4$ a small deviation 
from the SU(4)
point is irrelevant\cite{azaria1}, hence the low energy properties of the model is still
controlled by the SU(4)$_1$ fixed point mentioned above. In contrast, for $J>K/4$, the
deviation results in marginally relevant interactions which open a gap in the spectrum, and the ground
states are dimerized with alternating spin and orbital singlets. The low-lying excitations are just the fermions of the SO(6) Gross-Neveu model. 

Recently, the region where $J_1\neq J_2$
has been explored in Ref.\onlinecite{affleck2,azaria2}, and it was shown that the gapless phase can be extended to a 
large region around the original gapless line. (region V in Ref.\onlinecite{affleck2} or 
phase B in Ref.\onlinecite{azaria2}). Moreover, the low energy physics is 
described by SU(2)${_{2,s}\otimes}$SU(2)$_{2,t}$ WZNW model, and the two level-two
SU(2) WZNW models are in general characterized by different velocities $u^*_s, u^*_t$.
These conclusions are also consistent with those obtained through numerical studies
\cite{udea2}.

On the other hand, the Kondo effect or more generally speaking, the quantum impurity
problem in one-dimensional strongly correlated electron systems is one of the central 
topics in condensed matter physics over the past few years. It is known that the 
interacting enviroment developed around each local  scattering center changes its character
drastically. For example, a weak potential scatterer renormalizes into an infinitely strong
blockade to transport\cite{fisher}, while a one-channel Kondo impurity develops properties 
reminiscent of the two-channel Kondo effect\cite{nagaosa}. From theoretical point of 
view, these problems usually provide interesting realizations of non-Fermi liquid physics, and the advances of nanofabrication techniques in the last few years make many of the 
above theoretical ideas can possibly be realized in laboratories. 

In this paper, we would like to study the effect of local imperfection and Kondo problem
of the one-dimensional spin-orbital model in its gapless phase. This can be viewd as a 
``stripped-down'' version of quantum impurity problems where the charge degrees of 
freedom have been projected out, and may correspond to the Hubbard model at special filling, which becomes 
insulating due to Umklapp scattering. 
The effects of magnetic impurities on 1D Heisenberg chain have already been studied by S. Eggert et al.\cite{affleck1},
and for the case of periodic chain, the {\it leading-correction-to-scaling boundary operator} (LCBO) is exactly the one that occurs in the two channel Kondo problem.
For our case, the presence of orbital degrees of freedom make us
expect nontrivial boundary critical behavior can occur through nontrivial LCBO. In fact, as we shall see, depending on the details, the system can flow to an open chain,
a periodic chain 
or an intermediate boundary fixed point, and nontrivial
scaling behavior can indeed occur. The content of this paper is organized as follows : In 
section II, we discuss bosonization formulas for bulk and open boundary conditions. Our results are presented in section III, and section IV is 
conclusions.

\section{Bosonization for bulk and boundary operators}

The model (\ref{ham1}) around the SU(4) point ($J_1\simeq J_2\simeq K/4$) can be bosonized 
from the SU(4) Hubbard model at quarter filling\cite{azaria1} :
\bea
H&=&\sum_i (-t c_{i+1 a\sigma}^{\dagger}c_{ia\sigma}+ H.c.) \nonumber \\
   &&+~\frac{U}{2} \sum_{iab\sigma\sigma^{'}} n_{ia\sigma}n_{ib\sigma^{'}}
   (1-\delta_{ab}\delta_{\sigma\sigma^{'}}) \ ,
\eea
by introducing the left and right movers for low energy degrees of freedom around the 
Fermi points ($k_F=\pi/{4a_0}$) :
\bea
\frac{c_{ia\sigma}}{\sqrt{a_o}}\simeq R_{a\sigma}(x)~exp(ik_Fx)
          +L_{a\sigma}(x)~exp(-ik_Fx) \ . \nonumber
\eea
At this point, we can bosonize the above slowly varying fields as usual  through 
introducing four chiral bosonic fields $\Phi_{a\sigma R/L}$ using the Abelian bosonization
formula\cite{affleck4} :
\bea
R_{a\sigma}&=&\frac{\kappa_{a\sigma}}{\sqrt{2\pi a_0}}
~exp(i\sqrt{4\pi}\Phi_{a\sigma R}) \ , \nonumber \\
L_{a\sigma}&=&\frac{\kappa_{a\sigma}}{\sqrt{2\pi a_0}}
~exp(-i\sqrt{4\pi}\Phi_{a\sigma L}) \ ,
\eea
where the bosonic fields satisfy the commutation relation $[\Phi_{a\sigma R},\Phi_
{b\sigma^{'} L}]=\frac{i}{4}\delta_{ab}\delta_{\sigma\sigma^{'}}$, and the Klein factors $\kappa_{a\sigma}$ introduced here are used to insure the anticommutation relations 
between different flavors of fermions, which satisfies the following anticommutation rule $\{\kappa_{a\sigma},\kappa_{b\sigma^{'}}\}=2\delta_{ab}\delta_{\sigma\sigma^{'}}$. 
The physical properties of the system can be made more transparent by changing to a new basis :
\bea
\Phi_c=\frac{1}{2}(\Phi_{1\uparrow}+\Phi_{1\downarrow}+\Phi_{2\uparrow}
            +\Phi_{2\downarrow}) \ , \nonumber \\
\Phi_s=\frac{1}{2}(\Phi_{1\uparrow}-\Phi_{1\downarrow}+\Phi_{2\uparrow}
            -\Phi_{2\downarrow}) \ , \nonumber \\
\Phi_f=\frac{1}{2}(\Phi_{1\uparrow}+\Phi_{1\downarrow}-\Phi_{2\uparrow}
            -\Phi_{2\downarrow}) \ , \nonumber \\
\Phi_{sf}=\frac{1}{2}(\Phi_{1\uparrow}-\Phi_{1\downarrow}-\Phi_{2\uparrow}
            +\Phi_{2\downarrow}) \ .  
\label{bosonic}
\eea

Umklapp scatterings arising at higher order perturbation theory will result in a Mott  
transition at finite value of $U=U_c$, therefore for $U>>U_c$, the charge field $\Phi_c$ has 
a large gap, and only the spin-orbital part are left in the low energy sector. The remaining 
bosonized Hamiltonian can be further simplified by refermionization through the 
introduction of six Majorana fermions $\xi^a, a=1\ldots6$ :
\bea
(\xi^1+i\xi^2)_{R(L)}&=&\frac{\eta_1}{\sqrt{\pi a_0}}~exp(\pm i\sqrt{4\pi}\Phi_{sR(L)}) \ ,
\nn\\
(\xi^3+i\xi^4)_{R(L)}&=&\frac{\eta_2}{\sqrt{\pi a_0}}~exp(\pm i\sqrt{4\pi}\Phi_{fR(L)})\ , 
\nn\\
(\xi^5+i\xi^6)_{R(L)}&=&\frac{\eta_3}{\sqrt{\pi a_0}}~exp(\pm i\sqrt{4\pi}
\Phi_{sfR(L)}) \ , 
\eea
where $\eta_i$ are Klein factors.
The resulting Hamiltonian can then be written as\cite{azaria2} :
\bea 
{\cal H}&=&-\frac{i u_s}{2}({\mbox{\bm$\xi$}}_{sR}\partial_x
{\mbox{\bm$\xi$}}_{sR}-{\mbox{\bm$\xi$}}_{sL}
\partial_x {\mbox{\bm$\xi$}}_{sL})\nn\\
&&+~(G_1+G_3)(\kappa_1+\kappa_2+\kappa_6)^2\nn\\
&&-~\frac{i u_t}{2}({\mbox{\bm$\xi$}}_{tR}\partial_x{\mbox{\bm$\xi$}}_{tR}-
{\mbox{\bm$\xi$}}_{tL}\partial_x {\mbox{\bm$\xi$}}_{tL})\nn\\
&&+~(G_2+G_3)(\kappa_3+\kappa_4+\kappa_5)^2\nn\\
&&+~2G_3~(\kappa_1+\kappa_2+\kappa_6)(\kappa_3+\kappa_4+\kappa_5) \ ,
\label{ham2}
\eea
where the spin and orbital triplets are defined as : ${\mbox{\bm$\xi$}}_s=
(\xi^2, \xi^1, \xi^6)~\mbox{and}~{\mbox{\bm$\xi$}}_t=(\xi^4, \xi^3, \xi^5)$, $\kappa_a$ is defined as $\xi^a_R\xi^a_L$. 
$G_1~\mbox{and}~G_2$ measure the deviation from the SU(4) point, i.e. $J_1=\frac{K}{4}+G_1$
and $J_2=\frac{K}{4}+G_2$. $G_3<0$ is a nonuniversal parameter that could be extracted from the exact solution, and the two velocities $u_s, u_t$ are in general not 
equal to each other.


It was shown in Ref.\onlinecite{affleck2} and \onlinecite{azaria2} that the 
Hamintonian (\ref{ham2}) contains
several phases. Especially, there exist an extensive region where the system is gapless
and the the low energy fixed point is governed by a SU(2)$_{2,s}\otimes$SU(2)$_{2,t}$ WZNW model :
\bea
{\cal H}&=&{\frac{1}{2\pi}}\left\{\frac{u_s^*}{k_s+2}:{\bf J_s}\cdot{\bf J_s}:+
\frac{u_t^*}{k_t+2}:{\bf J_t}\cdot{\bf J_t}:\right\} \ ,
\label{bulkham}
\eea
where $k_s=k_t=2$, $u^*_s$ and $u^*_t$ are renormalized velocities of the fixed point 
theory for spin and orbital sectors, respectively. $J^a_{s,t}(x)~(a=1, 2, 3)$ are current 
operators for level two SU(2) WZNW model, whose Fourier modes obey the Kac-Moody algebra :
\bea
[J_n^a,J_m^b]&=&i\epsilon^{abc}J_{n+m}^c+\frac{1}{2}kn\delta_{n+m,0} \ . \nn
\eea
The spin and orbital density operators have the following general forms:
\bea
{\vec S}_i&\sim&{\vec J}_{sR}+{\vec J}_{sL}+~\left(e^{i\pi x/2a_0}{\vec{\cal N}}_s+
\mbox{H.c.}\right)\nn\\  &&+~(-1)^{x/a_0}{\vec n}_s ,\nn\\
{\vec T}_i&\sim&{\vec J}_{tR}+{\vec J}_{tL}+~\left(e^{i\pi x/2a_0}{\vec{\cal N}}_t+
\mbox{H.c.}\right)\nn\\ &&+~(-1)^{x/a_0}{\vec n}_t ,
\eea
here ${\vec J}_{s,t}$ are the smooth $(k\sim 0)$ parts of the spin( orbital) density, while
${\vec{\cal N}}_{s,t}$ and ${\vec n}_{s,t}$ are the $2k_F=\pi/2a_0$ and $4k_F=\pi/a_0$
parts. 

The current operators can be expressed in terms of Majorana fermions :
\bea
{\vec J}_{sR(L)}&=&-\frac{i}{2}{\mbox{\bm$\xi$}}_{sR(L)}\wedge
{\mbox{\bm$\xi$}}_{sR(L)} \ ,\nn\\
{\vec J}_{tR(L)}&=&-\frac{i}{2}{\mbox{\bm$\xi$}}_{tR(L)}\wedge
{\mbox{\bm$\xi$}}_{tR(L)} \ .
\label{current2}
\eea
The boson representations for $2k_F$ components ${\vec{\cal N}}_{s,t}$ are :
\bea
{\cal N}_s^z&\propto&exp~\{i\sqrt{\pi}(\Phi_s+\Phi_f+\Phi_{sf})\}\nn\\
&&-~exp~\{i\sqrt{\pi}(-\Phi_s+\Phi_f-\Phi_{sf})\}\nn\\
&&+~exp~\{i\sqrt{\pi}(\Phi_s-\Phi_f-\Phi_{sf})\}\nn\\
&&-~exp~\{i\sqrt{\pi}(-\Phi_s-\Phi_f+\Phi_{sf})\} \ , \nn\\
{\cal N}_t^z&\propto&exp~\{i\sqrt{\pi}(\Phi_s+\Phi_f+\Phi_{sf})\}\nn\\
&&-~exp~\{i\sqrt{\pi}(\Phi_s-\Phi_f-\Phi_{sf})\}\nn\\
&&+~exp~\{i\sqrt{\pi}(-\Phi_s+\Phi_f-\Phi_{sf})\}\nn\\
&&-~exp~\{i\sqrt{\pi}(-\Phi_s-\Phi_f+\Phi_{sf})\} \ , \nn\\
{\cal N}_s^+&\propto&exp~\{i\sqrt{\pi}(\Theta_s+\Phi_f+\Theta_{sf})\}\nn\\
&&-~exp~\{i\sqrt{\pi}(\Theta_s-\Phi_f-\Theta_{sf})\} \ , \nn\\
{\cal N}_t^+&\propto&exp~\{i\sqrt{\pi}(\Phi_s+\Theta_f+\Theta_{sf})\}\nn\\
&&+~exp~\{i\sqrt{\pi}(-\Phi_s+\Theta_f-\Theta_{sf})\} \ ,
\label{2kf1}
\eea
where $\Theta_a$ are the dual fields of $\Phi_a$, and satisfy $[\Phi_a(x), \Theta_b(y)]=
i\delta_{ab}\Theta (y-x)$.

The $2k_F$ components can be written in a more compact way by noting that the six
Majorana fermions could be associated with six critical Ising models. Then using the order and
disorder operators, $\sigma_a~\mbox{and}~\mu_a$, of the Ising models, they can be 
expressed as follows :
\bea 
{\cal N}^z_s&\propto&i\mu_1\mu_2\sigma_3\sigma_4\sigma_5\sigma_6
                                    +\sigma_1\sigma_2\mu_3\mu_4\mu_5\mu_6 \ , \nn\\
{\cal N}^z_t&\propto&i\sigma_1\sigma_2\mu_3\mu_4\sigma_5\sigma_6
                                    +\mu_1\mu_2\sigma_3\sigma_4\mu_5\mu_6  \ , \nn\\
{\cal N}^{+}_s&\propto&\left(\sigma_1\mu_2+i\mu_1\sigma_2\right)
     \left(\sigma_3\sigma_4\sigma_5\mu_6+\mu_3\mu_4\mu_5\sigma_6\right) \ , \nn\\
{\cal N}^{+}_t&\propto&\left(-\mu_3\sigma_4+i\sigma_3\mu_4\right)\left(\mu_1\mu_2
                      \sigma_5\mu_6-\sigma_1\sigma_2\mu_5\sigma_6\right) \ .
\label{2kf2}
\eea
The $4k_F$ part of spin and orbital operators, generated from higher harmonics of 
bosonization due to interactions, can be written down by noting that these operators should 
transform as vectors under SO(3)$_{s,t}$ and carry no chirality\cite{azaria1} :
\bea
{\vec n}_s&\propto&i{\mbox{\bm$\xi$}}_{sR}\wedge{\mbox{\bm$\xi$}}_{sL} \ , \nn\\
{\vec n}_t&\propto&i{\mbox{\bm$\xi$}}_{tR}\wedge{\mbox{\bm$\xi$}}_{tL} \ .
\label{4kf1}
\eea

Since the fixed point is governed by a SU(2)$_{2,s}\otimes$SU(2)$_{2,t}$ WZNW 
theory, it is better to write the above operators in a way which makes the 
symmetry properties
more transparent. This can be done by noting that each components of ${\vec S}_i$ 
and
${\vec T}_i$ should transform as a vector under 
spin and orbital SU(2) rotations,
respectively. This means that each components of ${\vec S}_i$ and ${\vec T}_i$ should
be primary fields of the SU(2)$_2$ WZNW model\cite{J}.
It can then be immediately seen that ${\vec{J}}_{s,t}$ are just the current operators of 
SU(2)$_{2s,t}$ WZNW models, and the $2k_F$ components correspond to the 
spin $1/2$
primary fields of it. The latter can be made evident from Eq. (\ref{2kf2}) by using the equivalence between
a SU(2)$_2$ WZNW theory and three critical Ising models\cite{zamo}. Especially, the 
spin $1/2$ 
primary field can be expressed as the product of three order or disorder operators of the
corresponding three Ising models. We then have :
\bea
{\cal N}^a_s&\sim&g^{(2)}_{sa}g^{(2)}_{t0}-ig^{(1)}_{sa}g^{(1)}_{to} \ , \nn\\ 
{\cal N}^a_t&\sim&g^{(2)}_{s0}g^{(2)}_{ta}-ig^{(1)}_{s0}g^{(1)}_{ta} \ ,
\label{2kf3}
\eea
where $a=1, 2, 3$ and $g^{(1,2)}_{s,t\alpha}$ are defined as: 
\bea
g&=&\tau^{\alpha}\!\!\left(g_{\alpha}^{(1)}+ig_{\alpha}^{(2)}\right) \ . \nn
\eea
Here $\tau^{\alpha}$ are Pauli matrices for $\alpha=1, 2, 3$, $\tau^0$ is the identity matrix, and $g_{s,t}$ 
are spin one-half
primary fields of SU(2)$_{2s,t}$ WZNW theory. The remaining primary fields with 
spin one
$\Phi^{(1)}_{s,t}$ just correspond to the $4k_F$ components ${\vec n}_{s,t}$.

After completing discussions about the fixed point theory and its operator contents, we turn
to the bosonized forms of the above operators in open boundary condition. The open
chain boundary condition introduces the following boundary conditions on the left- and 
right- moving fermion fields\cite{affleck1} :
\[
R_{a\sigma}(0)+L_{a\sigma}(0)=0 \ , 
\]
when transformed into boson language, it becomes 
\[\Phi_{a\sigma R}(0)+\Phi_{a\sigma L}(0)=-\frac{\sqrt\pi}{2} \ . \]
We can then analytically continue the right-moving fields to left-moving fields by 
$\Phi_{a\sigma R}(x,t)=-\frac{\sqrt\pi}{2}-\Phi_{a\sigma L}(-x,t)$. In this way, we arrive
at a description of the system in terms of {\it chiral} fields only.

With the above relations, we find for the boundary fields, we have :
\bea
\Phi_{a\sigma}(x,t)&=&-\frac{\sqrt\pi}{2}+\Phi_{a\sigma L}(x,t)-\Phi_{a\sigma L}(-x,t)
\nn\\ &&\Rightarrow \Phi_{a\sigma}(0,t)=-\frac{\sqrt\pi}{2} \ , \nn\\
\Theta_{a\sigma}(x,t)&=&\frac{\sqrt\pi}{2}+\Phi_{a\sigma L}(x,t)+\Phi_{a\sigma L}
(-x,t)\nn\\ &&\Rightarrow \Theta_{a\sigma}(0,t)=\frac{\sqrt\pi}{2}+2\Phi_{a\sigma L}
(0,t) \ ,
\label{chiraleq1}
\eea
or in terms of Majorana fermions :
\be
 \xi^a_{R}(x,t)=\xi^a_{L}(-x,t) \ . 
\label{chiraleq2}
\ee
Substituting Eq. (\ref{chiraleq1}), Eq. (\ref{chiraleq2}) into Eq. (\ref{current2}), 
Eq. (\ref{2kf1}) and Eq. (\ref{4kf1}), it is easy to see that all components of spin (orbital) operators are propotional to the current operators, i.e. :
\be
{\vec S}_{boundary}\propto {\vec J}_{sL}(0) \ , \ \ \ {\vec T}_{boundary}\propto {\vec J}_{tL}(0) \ .
\label{bundary}
\ee
This completes our discussions about the bosonization formulas.

\section{Boundary critical behavior}

In this section, we shall apply the bosonization formulas obtained in previous sections
to discuss the possible boundary fixed points. Two cases are considered here : local defects
which result in a change of local coupling strength compared with the bulk value
and an
external local moment coupled to the bulk system (Kondo impurity).

\subsection{Internal impurities}
As discussed by Eggert and Affleck\cite{affleck1}, in the case of local defect, there are 
two important
symmetries which distinguish the possible boundary fixed points of a Heisenberg chain 
: The site parity $P_S$ which
is reflection of the chain about one site, and the link parity $P_L$ which reflects the chain
about one link. Another important symmetry of the lattice system is the translation by one site $T$,
and we have the relation $P_L=P_S\otimes T$. As we shall see later, it is 
exactly the same symmetries which distinguish the possible boundary critical behaviors in
the case of spin-orbital model.

We first discuss the case where the local defect is invariant under link parity $P_L$, i.e.
altering the coupling strength of one link slightly. The corresponding operators are
${\vec S}_i\cdot{\vec S}_{i+1}$, ${\vec T}_i\cdot{\vec T}_{i+1}$, and
$({\vec S}_i\cdot{\vec S}_{i+1})\times({\vec T}_i\cdot{\vec T}_{i+1})$.
Using Eq. (\ref{2kf3}) and the following fusion rules for SU(2)$_2$ WZNW 
model\cite{ginzbarg} :
 \bea
&&g~\Phi\sim g \ , \nn\\
&&J^a_L(z)~g(\omega,\overline{\omega})\sim \frac{-\tau^a g}
{z-\omega} \ , \nn\\
&&J^a_R(\overline z)~g(\omega,\overline{\omega})\sim \frac{g \tau^a}
{\overline z-\overline \omega} \ ,
\label{fusionsu2}
\eea
where $g$ and ${\bf \Phi}$ are spin-$1/2$ and spin-$1$ primary fields, respectively.
It is then easy to see that the leading contribution from ${\vec S}_i\cdot{\vec S}_{i+1}$
and ${\vec T}_i\cdot{\vec T}_{i+1}$ is :
\be
{\hat O}_I=(i)^j (
const.~\tr g_s\cdot \tr g_t + const. ~\tr g^{\dagger}_s\cdot \tr g_t + \mbox{H.c.})\ . 
\label{boundaryop1}
\ee
Since both $g_s$ and $g_t$ have conformal dimensions $(3/16,3/16)$
, the above operator ${\hat O}_I$ has scaling dimension $\Delta=\frac{3}{4}$ and is
a relevant boundary operator. Hence a small deviation of the coupling will be drived to strong
coupling for either sign of $\delta J_{1,2}$. The remaining operator
$({\vec S}_i\cdot{\vec S}_{i+1})\times({\vec T}_i\cdot{\vec T}_{i+1})$ can be extracted
from fusion between two ${\hat O}_I$s and only contributes a marginal operator 
$\sim \tr {\bf \Phi}_s +\tr {\bf \Phi}_t$ which does not affect the RG flow.
At this point, one important difference between the one-dimensional spin-orbital model
and the usual antiferromagnetic Heisenberg chain should be noticed : For the spin-orbital model, 
the coupling constants between neighboring sites can be either 
antiferromagnetic (AF) or {\it ferromagnetic}\cite{affleck2,azaria2,udea2} (FM). With this in mind,
we expect the following possible strong coupling behaviors :
\begin{itemize}
\item Case I. $J_{1(2)}>0$ and $\delta J_{1(2)}>0$. The couplings will flow toward strong         AF
         coupling, and the local spin (orbital) degrees of freedom will form a {\it singlet}.
         The system becomes an open chain with two fewer spin (orbital) degrees of  
freedom.
         The stability of this strong coupling fixed point is guaranteed by the fact that the
         LCBO at this fixed point is just the product of two boundary spin (orbital) chiral 
        current operators which have dimension two, hence are irrelevant.
\item Case II. $J_{1(2)}<0$ and $\delta J_{1(2)}<0$. The couplings will flow toward strong FM
         couplings, and the local spin (orbital) degrees of freedom will form a {\it triplet}. 
         The
         system becomes an open chain with an additional triplet degree of freedom left.
         The stability of this strong coupling fixed point is guaranteed by the fact that the 
         residual coupling between the triplet moment and the open chain is 
         FM,
         hence is marginally irrelevant.
\item Case III. $J_{1(2)}>0$ and $\delta J_{1(2)}<0$ or $J_{1(2)}<0$ and 
         $\delta J_{1(2)}>0$. In this case,
         the coupling will flow to zero and leave a residual spin-orbital coupling $K
         ({\vec S}_0\cdot{\vec S}_1)({\vec T}_0\cdot{\vec T}_1)$. The fate of these 
degrees
        of freedom at impurity sites $0$ and $1$ depends on the details of combinations of various possibilities.
        For example, if spin $\in$ case I, orbital $\in$ case III, the two sites will first form
a spin singlet with a residual orbital degrees of freedom described by 
$\frac{-3}{4}K({\vec T}_0\cdot{\vec T}_1)$ which has a lower orbital triplet separated
from a higher orbital singlet by a gap $\sim O(K)$.  If in this case, $J_2<0$, then the strong coupling fixed point is 
just an open chain with an orbital triplet. If $J_2>0$, the triplet will be further screened
by neighboring sites due to Kondo screening and the strong coupling fixed point is just an 
open chain.
On the other hand, if spin $\in$ case II, orbital $\in$ 
case III, the two sites will form a spin triplet with a residual orbital degrees of freedom 
described by $\frac{1}{4}K({\vec T}_0\cdot{\vec T}_1)$ which has a ground state with an 
orbital singlet separated from the higher triplet state. Then with the same reasoning as 
previous 
discussion, one expect the strong coupling fixed point is either an open chain or an open
chain with a spin triplet.
In any case, they will still flow to either an open chain or an open chain 
with a residual ferromagnetic coupling to a spin or orbital triplet.
\end{itemize}
The resulting possible boundary critical behavior is summarized in table \ref{table1}.
Besides, we should mention that the appearance of a local triplet will not change the 
LCBOs, and only leads to an additional ground state degeneracy, hence an additional impurity entropy $S_{imp}=\ln 3$. Of course these asymptotically decoupled local moments
will also add a Curie-like contribution to ${\cal C}_{imp}$ and ${\cal \chi}_{imp}$, in
additional to logarithmic corrections characteristic of asymptotic freedom.

We now turn to the case where the local defect respects site parity $P_s$, i.e. varying the 
coupling strength of two adjacent links by the same amounts. In this case, the leading 
boundary operator arises from the sum of ${\vec S}_i\cdot{\vec S}_{i+1}$ and 
${\vec T}_i\cdot{\vec T}_{i+1}$ between two adjacent links. Due to the staggering factor  in front of the $2k_F$ and $4k_F$ components, in the continuum limit, it is just the 
differential of 
Eq. (\ref{boundaryop1}) :
\be
\frac{d}{dx}\left(const.~\tr g_s\cdot \tr g_t + const. ~\tr g^{\dagger}_s\cdot \tr g_t + 
\mbox{H.c.}
\right) \ ,
\label{boundaryop2}
\ee
which has scaling dimension $1+\frac{3}{4}=\frac{7}{4}$. Therefore, we conclude that a small
deviation  of coupling strength of two adjacent sites is irrelevant and the low energy fixed point
is just a periodic chain. 
Since the open chain with a decoupled spin or orbital doublet is 
stable only for ferromagnetic couplings, we arrive at the conclusion that when $J_{1,2}>0$, the open
chain will be unstable and flow to the stable periodic chain with the impurity site included.
However, for $J_{1,2}<0$, the open chain and periodic chain fixed points are not 
connected by a monotonous RG flow.

From the above discussions, we find that for a local defect, there can be two
possible boundary critial behaviors similar to that of a Heisenberg chain\cite{affleck1} :
\begin{itemize}
\item For impurities which violate the site parity $P_s$ , the infrared fixed point corresponds
to an open boundary condition.
\item For impurities which respect $P_s$ (hence violate $P_L$), the local defect is 
irrelevant if the deviation of coupling strength is not too large and
at low energy, the chain ``heals''.
\end{itemize}
\noindent
\begin{minipage}{0.48\textwidth}
\begin{table}
\caption{Possible boundary critical behaviors for a local defect respect link parity $P_L$
(OC represents the open chain, $T_s$ and $T_t$ represent residual spin and orbital
triplets, respectively) :}
\begin{tabular}{p{2.5cm}cccc}
& $J_1>0$ & $J_1>0$ & $J_1<0$ & $J_1<0$  \\
& $J_2>0$ & $J_2<0$ & $J_2>0$ & $J_2<0$  \\ 
\hline
OC&$\surd$ &$\surd$ &$\surd$  &   $\surd$ \\ 
OC$+T_s$&$\surd$ &$ $ &$ \surd$ &$\surd  $ \\ 
OC$+T_t$&$\surd$ &$ \surd$& $ $& $\surd $   \\ 
OC$+T_s+T_t$& &$ $ & &$\surd$  \\
\end{tabular}
\label{table1}
\end{table}
\end{minipage}

\subsection{External impurities}
In this section, we discuss the boundary critical behavior for an external spin $1/2$ local 
moment coupled to the bulk system via AF exchange (Kondo) coupling.
The Hamiltonian is decomposed as ${\cal H}={\cal H}_0+{\cal H}_K$ where 
${\cal H}_0$ is Eq. (\ref{bulkham}) and ${\cal H}_K$ is :
\bea
{\cal H}_K&=&J_K {\vec S}_{imp}\cdot {\vec S}_0  \ ,  \nn\\
&=& {\vec S}_{imp}\cdot \left\{\lambda_F{\vec J}_{s}(0)+
\lambda_{B}({\vec N}_s(0)+{\vec N}_s^{\dagger}(0))\right. \nn\\
&&\left.+~\lambda_{4k_F}{\vec n}_{s}(0) \right\} \ . \ 
\eea
Here $J_K>0$ is the Kondo coupling. $\lambda_F, \lambda_{B}, \lambda_{4k_F}$ are 
the forward, backward , $4k_F$ scattering strength, respectively, and all are proportional
to $J_K$. Since the Kondo coupling is 
antiferromagnetic, all the above coupling constants are relevant. In fact, they satisfy the 
renormalization group equations :
\bea
\frac{d \lambda_F}{d \ln L}&=&\frac{1}{2\pi u^*_s}(\lambda^2_F + 
\lambda_{4k_F}^2) \ , \nn\\
\frac{d \lambda_{B}}{d \ln L}&=&\frac{1}{4}\lambda_B +
{\cal O}(\frac{\lambda_B \lambda_F}{\pi u^*_s}) \ , \nn\\
\frac{d \lambda_{4k_F}}{d \ln L}&=&\frac{1}{\pi u^*_s}\lambda_F\cdot 
\lambda_{4k_F}  \ . 
\label{RGeq}
\eea
Note that the most crucial difference between forward scattering and other couplings is that
the latter breaks chiral symmetry and they couple orbital degrees of freedom to the 
impurity spin. Also note that backward scattering is the most 
relevant one which will flow to strong coupling for both ferro- and antiferromagnetic 
couplings. For the conveniences of latter boundary conformal field analysis, we should first
transform this problem into a chiral form : 
Both the spin and orbital sectors of Hamiltonian (\ref{bulkham}) can be mapped to a 
chiral theory by first folding the system to the positive $x-$axis through defining
${\vec J}^1_{L/R}(x)\equiv{\vec J}_{L/R}(x)$, ${\vec J}^2_{L/R}(x)\equiv
{\vec J}_{R/L}(-x)$. Interpreting the time axis as a boundary where  ${\vec J}^1_{L/R}
(\tau,0)={\vec J}^2_{R/L}(\tau,0)$, we can then analytically continue the theory back to 
the whole $x-$axis and it is described by two chiral (left-handed) currents.
Furthermore, since backward scattering breaks the SU(2)$\otimes$SU(2) symmetry down to its diagonal
one, it is better to write the Hamiltonian in a way which makes the symmetry more
transparent. For this purpose, we use the following {\it coset} construction\cite{ginzbarg} :
\bea
SU(2)_2\otimes SU(2)_2\sim SU(2)_4\otimes {\cal G}\ ,
\label{coset0}
\eea
where ${\cal G}$ is the $N=1$ SUSY unitary minimal model with $c=1$. By 
matching the
scaling dimensions and spin properties, we can establish the relations between the product
of conformal towers in these two representations. For our present purpose, we only need the following :
\bea
\left(\frac{1}{2}\right)_2 \times \left(\frac{1}{2}\right)_2 = (0)_4 \times 
\left[\phi_{(2,1)}\right] + ({\bf 1})_4\times \left[\phi_{(2,3)}\right]\ ,
\label{coset}
\eea
where $[\phi_{(p,q)}]$ are conformal towers of ${\cal G}$. The 
corresponding primary fields have conformal
dimensions $h_{p,q}=\frac{(3p-2q)^2-1}{48}+\frac{1}{32}[1-(-1)^{p-q}]$. $(j)_k$
is the conformal tower of SU(2)$_k$ WZNW theory with spin $j$~.
With this new representation, the spin (orbital) current operator 
${\vec J}_s(0)={\vec J}_{sL}(0)+{\vec J}_{sR}(0)={\vec J}^1(0)+{\vec J}^2(0)\equiv
{\bf J}(0)$ is now the current operator of chiral SU(2)$_4$ WZNW model, and
the Kondo interaction can be rewritten as\cite{4kf} :
\bea
{\cal H}_K=\left[ \lambda_F {\bf J}(0) + {\tilde\lambda}_B {\bf \Phi}(0)\phi^s_{(2,3)}
\phi^t_{(2,1)}\right]\cdot {\bf S}_{imp} \ ,
\label{kondoint}
\eea
where ${\tilde\lambda}_B$ is propotional to $\lambda_B$, and ${\bf \Phi}$
is the spin one primary field of SU(2)$_4$ WZNW theory. It can then be immediately seen that if both $\lambda_B$ and $\lambda_{4k_F}$ vanish, this problem can be solved as the
usual Kondo problems\cite{frojdh,affleck5} : 
The impurity spin can be absorbed at special value of coupling 
$\lambda^*_F=2{u^*_s}/{(2+k)|_{k=4}}$ by redefining the spin current as the sum of 
that of electrons and
impurity, ${\bf J}(x)+2\pi {\bf S}_{imp}\delta (x)\rightarrow {\bf J}(x)$. The 
boundary fixed point in this case is known to be that of the {\it four-channel} Kondo 
problem 
whose LCBO is the Kac-Moody descendant of the spin one primary field ${\bf J}_{-1}
\cdot {\bf \Phi}^{(1)}$ which leads to non-Fermi-liquid corrections to thermodynamical
quantities. However, as we shall argue in
the following, backward scattering will destabilize the four-channel Kondo 
intermediate fixed point. The system will be drived to the strong coupling open chain fixed
point or a new intermediate fixed point depending on whether the coupling strength of
orbital sector $J_2$ is ferro- or antiferromagnetic.

In fact, from eq. (\ref{RGeq}), we expect that $\lambda_B$ will scale to infinity first
and dominate the low energy physics. To get a physical picture, we can do strong coupling analysis of the original lattice model : At strong coupling, the impurity spin forms a singlet with the spin at site zero and leaves an orbital doublet behaind. For the case where 
$J_2<0$, the stability of this fixed point is guaranteed. We therefore conclude that in this case, the boundary fixed point is just an open chain. In other word,
{\it no} LCBOs can come from the s=1 conformal tower of the spin sector. The leading 
operators are then the usual dimension two operators 
${\bf J}^2$ which originate from exchange interaction between edge spins and orbitals.
However, in the case where $J_2>0$, the strong coupling fixed point is unstable and
the system will flow to some unknown intermediate fixed point. 
To determine the fate of the RG flow, we consider a special point in our parameter space 
with $J_2>0$, i.e. the SU(4) point where the underlying microscopic Hamiltonian can be
considered as the SU(4) Hubbard model. Away from quarter filling, the Hilbert space
can be decomposed into three parts : charge-, spin- and flavor conformal towers, and 
since the Kondo interaction breaks chiral invariance, operators with nonzero charge, spin
and flavor quantum numbers can occur as long as they transform as singlets under
the corresponding daigonal subgroups. The possible impurity critical behavior can be 
determined by identifying LCBOs which satisfy the following two conditions
\cite{frojdh,johannesson} : (i) {\it produce a noninteracting limit consistent with known 
results}, and (ii) {\it respect the symmetries of the Hamiltonian}. 
By transforming to a basis with definite parity, it is easy to see that only
channels with positive parity coupled to the impurity spin, hence the noninteracting limit of SU(4) Hubbard model corresponds to two-channel Kondo fixed point. Using this as a
reference point, we can borrow the results of Ref.\onlinecite{johannesson} : The two 
leading boundary operators are $e^{{i\sqrt\pi}/{4K_\rho}\phi_c}\times\phi_s \times\phi_f$
and ${\bf J}_{-1}^1\cdot{\bf \Phi}^1+{\bf J}_{-1}^2\cdot{\bf\Phi}^2$, where $K_\rho$ is
the Luttinger liquid paremeter and $\phi_c$ is the charge field. $\phi_{s,f}$ are singlet fields 
under the diagonal SU(2) subgroup and they are equal to $\phi_{(2,1)}^{s,t}$
in terms of our previous coset language. Upon approaching quartering, Umklapp scattering
opens a charge gap for $\phi_c$ and the remaining part of the first boundary operator has
dimension smaller than one, therefore should be surpressed by selection rules. Only the 
second interaction independent boundary operator survives in our case. Since we do not 
expect any qualitative change when the parameters deviate from the SU(4) point slightly as long as we
are still in the same phase, together with
the above renormalization group analysis we conclude that the system flows to a two channel Kondo fixed point for antiferromagnetic orbital 
couplings.

\subsection{Thermodynamical behavior}
In this section, we briefly discuss the corrections of specific heat and magnetic 
susceptibility induced by LCBOs, i.e. the corrections of LCBO $\lambda_I {\cal O}(0)$ to
the fixed point theory ${\cal H}^*$. The first type of LCBOs corresponding to open chain 
fixed
point are the Virasoro decendants of identitiy operators : ${\bf J}_1^2$, 
${\bf J}_2^2$, and ${\bf J}_1\cdot{\bf J}_2$. These operators will contribute to impurity
free energy defined by $\delta f_{imp}(T,\lambda_I)=f_{imp}(T,\lambda_I)-f_{imp}
(T,0)$ 
in the {\it first order} of $\lambda_I$. Since these operators have dimension two, we 
expect that $\lambda_I$ is proportional to $1/T_K$, where $T_K$ is the temperature scale at 
which the coupling strength of the relevant perturbations becomes of order one. Because
the relevant operators are of dimension $3/4$, we 
have $T_K\propto {\delta J}^4/v$ for a small amount of initial change of coupling strength $\delta J$ .
Then from dimensional 
consideration, we expect the correction to the specific heat ${\cal C}_{imp}$ should 
be proportional to $T/T_K$, and for the same reason, the correction to the magnetic 
susceptibility $\chi_{imp}$ should be proportional to $1/T_K$, i.e. at $T\rightarrow 0$,
it produces a $T$ independent behavior. Note that this result is identical to that of the
Heisenberg chain except the ``Kondo temperature'' $T_K$ has a different scaling 
relation with the coupling strength.

The second type of LCBO corresponding to periodic chain boundary condition is 
Eq. (\ref{boundaryop2}) which as we shall see, is 
a Virasoro primary operator. Consequently, its finite-temperature expection value vanishes,
and its contribution to $f_{imp}$ only starts from the {\it second order} of
$\lambda_I$. In order to proceed the caculations, it is better to transform into a chiral 
representation as previous subsection. To do this, we first note that 
Eq. (\ref{boundaryop2})  can be written as
\(
{\bf J}_{-1,s}\cdot {\mbox{tr}g_s{\mbox{\bm$\sigma$}}}{\mbox{tr}g_t}+
{\mbox{tr}g_s}
{\bf J}_{-1,t}\cdot {\mbox{tr}g_t{\mbox{\bm$\sigma$}}}
\), where ${\bf J}_{-1}={\bf J}_{R,-1}+{\bf J}_{L,-1}$ and is equal to 
${\bf J}_{1,-1}+{\bf J}_{2,-1}$ in terms of chiral fields. At this point, we 
can use the coset relation (\ref{coset0}) to cast the above operator into a simpler form. In 
fact, noting that $\mbox{tr}g{\mbox{\bm$\sigma$}}\rightarrow 
{g^{\alpha}_1{\mbox{\bm$\sigma$}}^
{\beta}_{\alpha}}g_{2\beta}\rightarrow{\bf\Phi}\phi_{(2,3)}$ and $\mbox{tr}g\rightarrow 
g^{\alpha}_1 g_{2\alpha}\rightarrow\phi_{(2,1)}$, the leading irrelevant operator
becomes ${\bf J}_{s,-1}\cdot{\bf\Phi}_{s}\phi_{(2,3)}^s\phi_{(2,1)}^t+(s\leftrightarrow t)$.
Following methods in Ref.\onlinecite{affleck5}, it can be shown that second-order 
perturbation theory results in an impurity specific heat :
\bea
{\cal C}_{imp}&=&\lambda_I^2 A\left(\frac{1}{u_s^{*2\Delta-3/4}u_t^{*3/4}}
+\frac{1}{u_t^{*2\Delta-3/4}u_s^{*3/4}}\right)\nn\\&&\times~\frac{2\Delta}
{3(2\Delta-3)}{\pi}^2 \tau_0^{3-2\Delta}T \ ,\nn
\eea
where $\Delta=\frac{7}{4}$ is the dimension of $\frac{d}{dx}{\hat O}_I$, $\tau_0$ is an 
infrared cutoff,
and $A=3(2+\frac{k}{2})=12$. Similarly, the impurity magnetic susceptibility can be 
obtained :
\bea
{\cal \chi}_{imp}=\frac{\lambda_I^2 A^{\prime}}{u_s^{*2\Delta-3/4}u_t^{*3/4}} 
{\tau}_0^{3-2\Delta}+O({\sqrt T}),\nn
\eea
where $A^{\prime}=2(2+\frac{k}{2})^2=32$. Note that although this LCBO looks 
nontrival, it only produces Fermi-liquid like behaviors because its dimension is
``too high''.

The third type of LCBO is the Kac-Moody descendent of spin one primary field from the 
spin and orbital sector : ${\bf J}_{-1}^1\cdot {\bf \Phi}^1+{\bf J}_{-1}^2\cdot 
{\bf \Phi}^2$ which appears in the case of Kondo impurity
with antiferromagnetic orbital couplings and is equal to 
${\bf J}_{-1}\cdot{\Phi}\phi_{(3,3)}$ in terms of coset representation, where ${\bf J}$
and ${\Phi}$ are now elements of $k=4$ WZNW theory . It is exactly the same leading 
irrelevent operator as the {\it two-channel} Kondo problem except now it involves both
spin and orbital sectors. It produces the following well-known form of impurity specific 
heat and magnetic susceptibility\cite{affleck5} :
\bea
{\cal C}_{imp}\sim \lambda^2_I T \ln {(T_k/T)}\ , \, \, 
{\cal \chi}_{imp}\sim \lambda^2_I \ln {(T_k/T)}\, . \nn
\eea

\section{Conclusions and discussions}
To summarize, we have studied possible boundary critical behaviors for the 
one-dimensional spin-orbital model in its gapless phase with a magnetic impurity. For the case of internal 
impurities, there can be either an open chain or periodic chain fixed point. The underlying
reason for the occurrence of these two different critical behaviors is similar to that of the
Heisenberg spin chain : The leading instability for periodic chain is determined by the 
spin and orbital dimerization operators $\epsilon_s (x)=i^j ({\vec S}_j\cdot
{\vec S}_{j+1})$ and $\epsilon_t (x)=i^j ({\vec T}_j\cdot{\vec T}_{j+1})$ . Although  they
are allowed for impurities which violate the site parity $P_s$, they are prohibited for 
impurities respecting $P_s$, with $\partial_x \epsilon_{s,t}(0)$ being the leading irrelevant
operators. The new feature in this case is that due to the existance of additional orbital
degrees of freedom and that the couplings between sites can be ferromagnetic, there can
be spin or (and) orbital triplet at the impurity site together with the open chain at the low energy fixed point.

For the case of Kondo impurity, we see that it can either flow to an open chain fixed point
or an intermediate fixed point depending on the sign of orbital couplings.
This should be compared with the case of Kondo effect in 
Luttinger liquid or Hubbard model at incommensurate filling. In that case, the backward
scattering which breaks chiral symmetry will result in nontrivial leading irrelevant 
operators from the charge sector with Q$_R-$Q$_L\neq 0$. The operators arising from 
coset construction
obtained via diagonal embedding (SU(2)$_{R}\otimes$SU(2)$_L /$SU(2)$_{diag}$) can 
also appear together with the ones from the charge sector through nontrivial {\it selection rules}. In fact, these leading irrelevant operators correspond to electron hopping with spin 
(orbital) flip between two ends in the open chain fixed point. In the present case, 
since the charge field is gapped, these processes cannot occur and only exchange 
interaction between edge spins (orbitals) can appear as LCBOs. A new feature in this case
is that when orbital coupling is AF, the strong coupling fixed point will be destabilized and
thermodynamical quantities at the intermediate fixed point show interesting temperature
dependence similar to that of the two-channel Kondo effect.

Finally, since our theory is characterized by two different velocities, and in the case of
periodic chain boundary fixed point, the contribution to ${\cal C}_{imp}$ and 
${\cal \chi}_{imp}$ from the LCBO is the same as that of the dimension two Virasoro 
descendents ${\bf J}^2$, a universal Wilson ratio can not be defined generally.

\acknowledgements

This work was supported by National Science Council of R.O.C. under the Grant No.
NSC89-2811-M-007-0015.

\end{multicols}


\begin{thebibliography}{99}
\bibitem{J}The only exception is the current operators which are of course not primary operators. However, the SU(2)
generator is $\int J^a(x)\sim J^a_0$, and we have $[J^a_0,J^a_n]=i\epsilon^{abc}J^c_n$.
\bibitem{4kf} Here we neglect the $4k_F$ part because its symmetry properties are similar
to thoses of the backward scattering part and it has higher dimensions. This does not affect 
our conclusions.
\end{thebibliography}
\end{document}